# Human-like general language processing

**Feng Qi 1,2** [*]
1. Intelligent Innovation Lab
Alibaba Group
Beijing, 100101
`qianlong.qf@alibaba-inc.com`

**Guanjun Jiang 1**
2. Nuffield Department of Clinical Neurosciences
University of Oxford
Oxford, OX39DU
`guanj.jiangjg@alibaba-inc.com`

## Abstract

Using language makes human beings surpass animals in wisdom. To let machines understand, learn, and use language flexibly, we propose a human-like general language processing (HGLP) architecture, which contains sensorimotor, association, and cognitive systems. The HGLP network learns from easy to hard like a child, understands word meaning by coactivating multimodal neurons, comprehends and generates sentences by real-time constructing a mental world model, and can express the whole thinking process verbally. HGLP rapidly learned 10+ different tasks including object recognition, sentence comprehension, imagination, attention control, query, inference, motion judgement, mixed arithmetic operation, digit tracing and writing, and human-like iterative thinking process guided by language. Language in the HGLP framework is not matching nor correlation statistics, but a script that can describe and control the imagination.

## 1 Introduction

Future strong machine intelligence requires intelligent language processing techniques. We believe that humans have such ability but modern machines don't, which can be reflected in, but not limited to, the following aspects. First, learning is a step by step process. The human brain can gradually assimilate and accumulate various concepts, knowledge and skills. But, current natural language processing (NLP) machine does not seem to care about the order of learning, rather than the quantity of corpus materials that contributes to robust relational statistics among words [1]. Second, word meaning is perceived by virtue of multimodal neuronal activation. For example, when we hear the word 'acid', we can feel sour as the saliva is excreted due to the activation of our gustatory neurons. However, if you ask NLP machine what is acid, ideally, it queries out the dictionary explanation of '[n] a chemical substance that neutralizes alkalis or [adj] having a pH value of less than 7', but we know that NLP itself does not know what the words 'neutralize', 'alkalis' or 'pH value' mean. Third, humans comprehend and generate sentences by real-time constructing a virtual world model. For example, when I said 'I have a gift in the box', you may naturally think of a piece of chocolate or ring being placed in the box. Based on the imagined scenario, you may ask 'is it chocolate?'. On the contrary, NLP focuses on the embedding matching between question and answer, and the correlation score for output [2]. So, NLP knows 'Donald Trump' and 'US President' are strongly linked, but does not imagine it as 'an old white man with shining blonde hair sits in the Oval Office'. Fourth, human thinking is a language guided process that is consciously describable and self-controllable. For example, a child can distinguish monkeys from humans and explain the judgment by saying 'monkeys have tails, but humans don't'. However, the current NLP only be able to report the classification results mechanically, but never be aware of its own thinking process, let alone control the thinking process [1-4]. Finally, humans can understand and apply language command at one trial. For example, we can play Gomoku once after knowing the rule sentence of 'win if Five in a row'. However, modern machines cannot understand the rule verbally and has to be trained billions of times with reinforcement learning merely for one such skill [4].

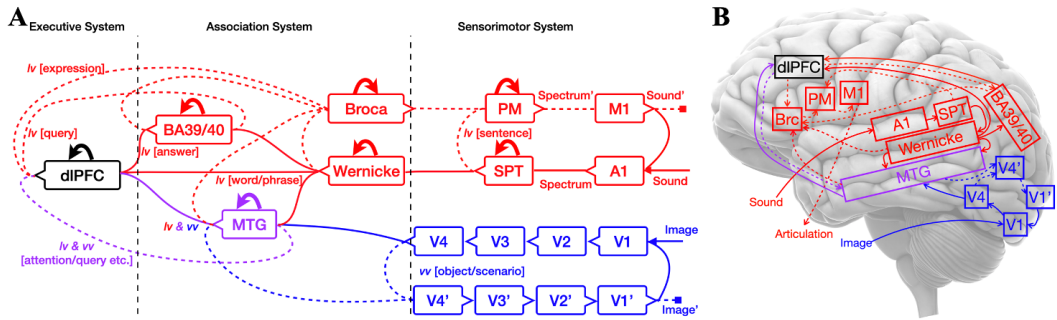

Figure 1: The architecture of the HGLP. (A) It consists of three hierarchies of sensorimotor, association and executive systems. The low-level sensorimotor cortices are made by visual and language autoencoders which are trained with unsupervised learning in the early stage. The post-trained visual and language autoencoders could provide visual vector (*vv*) of images viewed and language vector (*lv*) of sound heard, respectively. In the association cortices, there are middle temporal gyri (MTG) and intraparietal lobe (IPL, BA39/40), functioning as *lv-vv* translator and *lv-lv* associator, respectively. Wernicke area comprehends a sentence by decomposing it into words or phrases, while the Broca area constructs a sentence with various inputs of words and phrases syntactically. The high-level dorsal lateral prefrontal cortex (dlPFC) acquires tasks and environmental states by receiving the *lv* and *vv* vectors. It keeps the information in the working memory, generates task response according to rules, and top-down control signals to lower level modules to properly interact with the environment. (B) Anatomical connection of HGLP modules in human brain. Red represents language paths, blue represents visual paths, purple represents both modalities; solid represents feedforward paths, and dashed represents feedback paths.

Solving these problems will bring us closer to strong machine intelligence. Based on the research of the human brain neuroscience, we propose a human-like general language processing (HGLP) architecture, which aims to build a new language processing architecture through mimicking the brain with correct function implementation of cortical modules and information interaction among these modules.

## 2 HGLP Achitecture

Machine language processing could follow the human brain blue-print [5, 6] in order to achieve human-like general language processing skills. To construct such language processing architecture with sufficient simplicity, it is necessary to guarantee that the connections among cortical modules are correct, the functionalities of each module are reasonable, and the training progress of each module is in a similar order to human development. Figure 1 demonstrates the scheme of HGLP architecture. In general, we follow Baddeley's model of working memory [7], which consists of a central executive system used to control two slave systems (sensorimotor cortices): the phonological loop (PL) and the visuospatial sketchpad (VSS). This is in line with people's perception and cognition. However, between the central executive and primary sensorimotor cortices, we add association cortices, such as middle temporal gyri (MTG) as a visual-language translator, and Intraparietal lobe (IPL or BA39/40) as a substrate of abstract knowledge. Next, we will talk about the neural functionalities and AI implementation of each module.

HGLP's sensorimotor system are implemented in the form of autoencoders, that can process visual and language inputs and generate imagination and articulation outputs. Human visual system develops before language system. Without language labels, visual cortices could develop its neural network under an unsupervised learning mechanism, namely that, imagine what have been viewed, and then adjust the neural network to make the imagination more consistent with the image viewed. For visual processing, we construct a visual autoencoder in Figure 1, with a visual encoder part (V1-V4) to extract a 32-byte visual vector (*vv*) from each image viewed, and a visual decoder part (V4'-V1') to reconstruct an image from a given *vv*. The *vv* can be either the untangled representation of viewed images for higher level processing [8] or top-down signals given by higher hierarchical modules to be imagined via the visual decoder. For example, MTG could pass the attention modulated *vv* to visual decoder for specific object imagination (Fig. 3F). For simplicity, we have not implemented the visual two-stream model yet, so the *vv* vector contains both object features and spatial information.

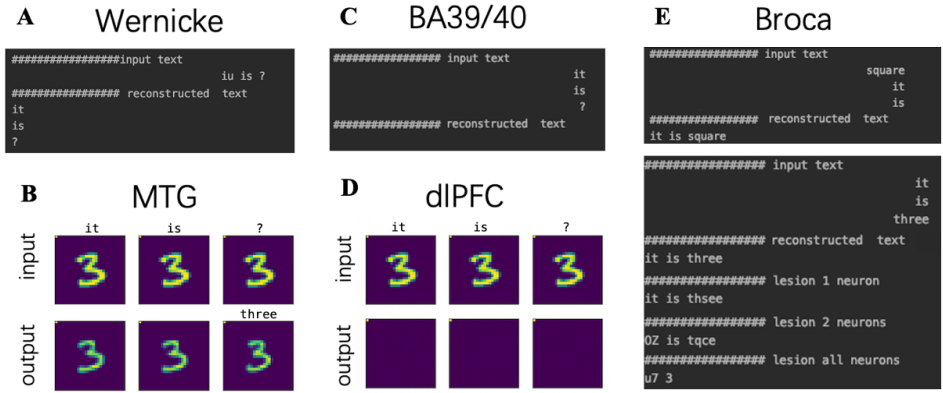

Figure 2: Object recognition task. (A) The HGLP visualized image '3' and heard 'it is ?', which were encoded into *vv* and *lv* by visual and language encoders. After that, Wernicke decomposed the sentence into word-level *lv*, also corrected the wrong pronunciation of 'iu' into 'it'. (B) MTG responded to the language command by identifying the digit and outputting the verbal identity 'three'. (C-D) BA39/40 and dlPFC did not respond to 'it is ?'. (E) Broca could combine *lv* from various modules grammatically for sentence utterance. Broca combined 'it is' from Wernicke and 'square' or 'three' from MTG into a sentence *lv* for future articulation via PM-M1. In the lesion test, Broca could still generate readable sentences with only 1 out of 32 neurons lesioned, its performance devastated rapidly if more than 2 out of 32 neurons were lesioned, and could only generate utterance 'u7 3' if all Broca neurons were silenced. The Wernicke, BA39/40 and Broca only process language-related *lv* vectors, while the MTG and dlPFC process both *lv* and *vv* vectors. The top of each block shows input image and language, and the bottom shows the reconstructed imagination and utterance according to the module output.

Babies learn to speak at one year old. After the auditory system processed the sound heard, the articulatory system could repeat such sound, and then the higher-level cortex could future process these language representations, such as endowing with meanings [9, 10]. HLGP also follows a similar architecture and processing work-flow. For language processing, we construct a language autoencoder, with a language encoder part (primary auditory cortex A1 and Sylvian parietal temporal (SPT)) [11] to convert the physical sound of a sentence into a 32-byte language vector (*lv*) and a language decoder part (vocal cord, laryngeal and tongue areas of PreMotor (PM) and M1) to articulate corresponding utterance from a given *lv*. Similarly, the *lv* can be either the language representation of heard sentence, or top-down signals given by higher hierarchical modules such as Broca for language generation. Here, the A1 module can convert physical sound into temporal spectrum signals, while M1 can articulate sound of the temporal spectrum signals (Supplementary Methods). The SPT-PM network is implemented by a sequence-to-sequence [12] model, where the PM aims to reconstruct accurately a temporal spectrum signal from a 32-byte *lv* encoded by SPT.

Human association cortex takes up a wide-spread cortical area between the sensorimotor cortex and the executive cortex. First, it is a knowledge center. Skills such as object recognition, motion detection are processed in MTG [13, 14], arithmetic computing in IPL [15], language comprehension in Wernicke area [16] and language generation in Broca area [16, 17], etc. Second, it is an information hub that receives multimodal representations processed by sensory cortices, and top-down query or control signals from executive cortices. After association processing, it provides reply according to high-level queries and signals to visual and language decoders for articulation and imagination. We implement the modules of the association system with LSTM [18] and use supervised learning to adjust the network parameters to acquire the corresponding functions.

Wernicke module is responsible for understanding the sound heard, mainly including the task of decomposing the utterance heard into phrases or words, which allows future semantic endowment by coactivating visual, gustatory and somatosensory, etc. neurons. To train Wernicke to have the sentence decomposition functionality, the language encoder A1-SPT is needed for the training data preparation by generating the *lv* of both input sentence and the expected output phrases and words. Wernicke can also filter out non-language sounds such as music and correct the external utterance according to the pronunciation and grammar rules. Therefore, the lesion of the Wernicke can influence the language comprehension [16]. Figure 2A demonstrates one example of Wernicke processing, which not only decomposed the input sentence into word-level *lv* but also corrected the wrong external pronunciation of 'iu' into 'it'.

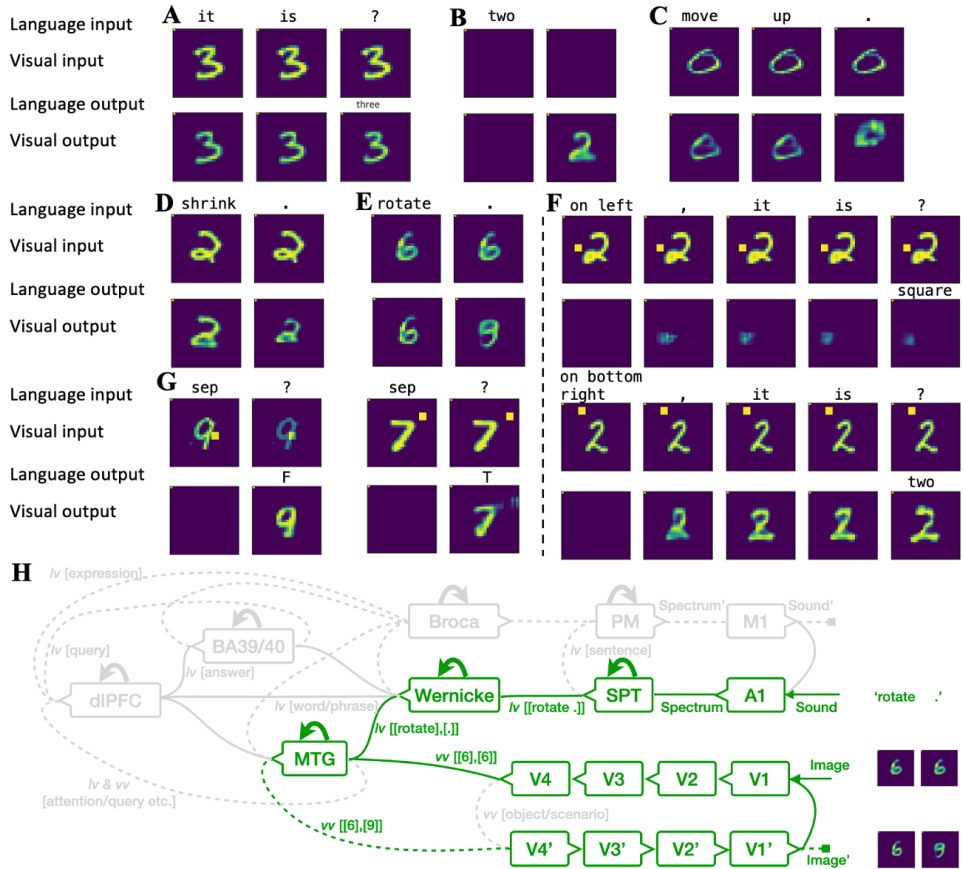

Figure 3: Language and visual interaction via MTG. (A) Recognition task: MTG could understand language commands and generate verbal identity according to the viewed object. (B) Imagination task: imagine digit according to language heard. (C) Imagined objects could be moved up, down, left and right according to language commands. (D) Language guided digit shrinking or enlarging. (E) Language guided digit rotating. (F) Language guided attention and object identification: MTG highlighted the object according to the heard prepositional phrase, and then identify it. (G) Motion prediction: MTG responded to command 'separated ?' and predicted whether the point will leave the digit at the next time step by outputting T or F *lv*. (H) Information flow of 'rotate' task. MTG converts *vv*[[6], [6]] into *vv*[[6], [9]] after hearing *lv* [[rotate], [.]]. Green indicates the activated path.

Broca module is responsible for the syntactic synthesis of languages (Fig. 2E), an opposite role of sentence decomposition of the Wernicke module. In general, our brain determines the sentence to be expressed before articulating it word by word. Broca plays a key role in synthetizing the sentence *lv* syntactically for language generation. Accordingly, we give Broca module in HGLP such functionalities via supervised training. Figure 2E demonstrated that Broca combines the 'square' from MTG and 'it is' from Wernicke into the sentence *lv* of "it is square", which could future be articulated by PM-M1. Therefore, Broca not only synthesizes sentences from various modules but also rearranged them according to the syntactical rules. Lesion to Broca does not affect language understanding but causes problems in language production, such as agramatical and effortful speech, namely, Broca aphasia [19]. Our lesion test in Figure 2E reveals a similar symptom. When a small number of neurons were silenced (1 out of 32 neurons was set to zero activation), the lesioned $lv$ could still be converted into a readable sentence; when a larger proportion of neurons (2/32 neurons) were lesioned, the language generation performance of Broca devastated rapidly; If most of the Broca neurons (32/32 neurons) were silenced, the articulating system (PM-M1) could only generate utterance 'u7 3', no matter what Broca intended to say, which showed quite similar symptoms to Broca's patient 'Tan' [20]. Also, since Broca's language generation is through contents synthesis, there is no exposure bias nor gradient vanishing issue in word by word language generation.

Human MTG is located between the language-related superior temporal cortex and visual-related inferior temporal cortex. It receives both visual and verbal processed information and functions as an interface or translator between these two modalities. The anterior MTG could explain a verbal name by co-activating the representation of associated visual neurons; on the other hand, after viewing an object, MTG could elicit activation of associated verbal neurons for naming articulation. Semantic dementia often has early neurodegeneration in this area [21], then patients could not tell names after seeing things, nor remember faces after hearing names. Posterior MTG is adjacent to parietal and occipital lobes, dealing with spatial and motion perception, which involves the association between visual scenario changes and verb understanding. We used supervised learning to give MTG such interactive functionalities between visual and language modalities. In the digit recognition task (Fig. 3A), MTG could understand the language command, process the $vv$ of viewed object accordingly, and generate the identity $lv$ that could finally be articulated via PM-M1. In the imagination task (Fig. 3B), MTG could imagine a digit according to language heard. In another word, MTG gives meanings to verbal words by coactivating corresponding visual neurons. By correctly manipulating imagined objects (Fig.3 C-E), we can claim that HGLP understands verbs and preposition such as move up/down/left/right, enlarge/shrink, rotate, etc. We also propose a language guided attention mechanism as is shown in Fig. 3F. After hearing the preposition phrases such as 'on left' or 'on bottom right', MTG shifted its attention to the corresponding objects for future processing, such as identity recognition. Such language-guided attention can also be given by the executive cortex via top-down control. Finally, Fig.3 G-H demonstrate that MTG could predict the visual motion of an object, and output whether point and digit will be separated. The 'subjective' judgment can be used in future tasks such as 'write and trace' in Fig. 4.

In addition, the human brain needs to understand abstract concepts and their relationships, such as the concept of 'even number', which cannot be explained by visual representation, but by verbal representation of 'number that can be divided by two is an even number'. The abstract information processing is approximately located at conjunction areas between parietal and temporal lobes. So, we construct a BA39/40 module to implement the corresponding functions. Fig. 4A shows parts of these functions, such as 8 is bigger than 4, 8 + 4 = 12 and F = ma, etc. During learning, these abstract relations are wired in the cerebral cortex as knowledge. When BA39/40 receives queries about such abstract knowledge (Fig. 4B), it can provide the answers accordingly.

The executive functions of the human brain are located in the prefrontal area [22, 23], which is considered to be orchestration of thoughts and actions to achieve internal goals. For simplicity, HGLP merely involves a dlPFC module with executive functionalities including task/rule recognition, attention, working memory, query, and inference. Human dlPFC is the neural substrate for task and rule identification and representation [24, 25]. In mixed arithmetic operation, the computing rule is ordered as parenthesis first, then multiplication and division, finally addition and subtraction, and left operation first at the same level. We constructed a dlPFC module trained on these arithmetic operations (four 1-digit numbers operations with addition, multiplication, and parenthesis). Figure 4B shows that the dlPFC could understand the language input of mathematical formula, correctly decompose the formula into single-step operations to be handled by BA39/40, save and retrieve the temporary results in the working memory, and generalize the rule to more numbers (five 1-digit numbers). For working memory, Fig. 4C demonstrates that the dlPFC could distinguish 'last' from 'this', by successfully presenting the previous digit with correct identity and shape without being affected by the current distractor. Moreover, the dlPFC could convert the verbal question 'what is ' to executable query command 'it is ?' for object identification that can be handled in MTG. Conditional selection 'if then' is the abstract expression of rule-based reasoning, which allows people to flexibly handle numerous tasks in real-time, such as syllogism reasoning. That is why all programming languages have 'if then' statement. We trained an LSTM to acquire such conditional selection (or abstract reasoning) capacity, with training data generated by template 'if *condition*, *action A*, else

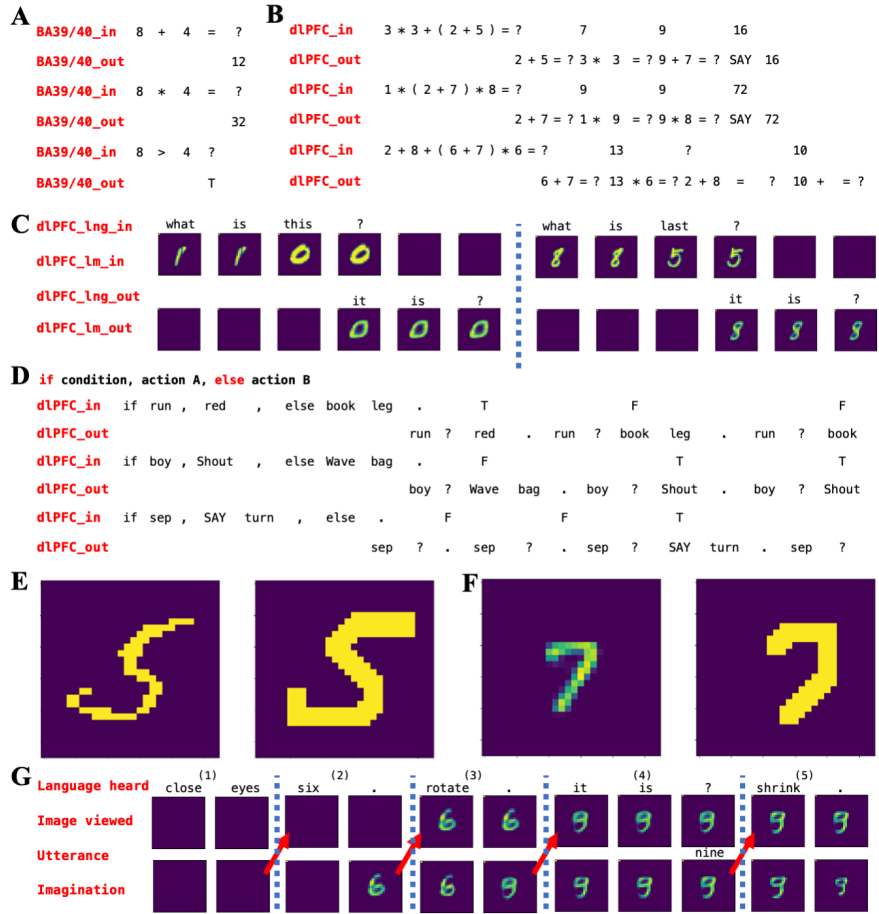

Figure 4: BA39/40 and dlPFC. (A) BA39/40 learnes and processes abstract knowledge such as arithmetic operation. (B) In mixed arithmetic operations, dlPFC is responsible for recognition of operation rules, while BA39/40 answers each single step operation. (C) dlPFC's role in working memory and task identification. It can distinguish 'this' from 'last', and convert question 'what is this/last ?' into query sentence 'it is ?' to trigger object identification in MTG. (D) dlPFC learned the conditional selection rule 'if *condition*, *action A*, else *action B*', where the training data of *condition*, *action A*, and *action B* are one or two randomly selected words from vocabulary. After hearing an 'if then' sentence, dlPFC outputs '*condition* ?' and the association cortex gives an answer of True or False, then dlPFC outputs *action A* or *B* respectively. (E) Word tracing task. After hearing the conditional selection sentence 'if sep, SAY turn, else ', the HGLP guided pen to trace the digit template. Left panel is the template and right is the traced result (the tracing process is displayed in supplementary movies). (F) Word writing task constitutes of imagining a digit and then tracing it. (G) Language guided iterative thinking process. (1) After closing the eyes, HGLP will not get any visual information from the outside world. (2) Language 'six .' is translated by MTG and the corresponding imagined figure is reconstructed by decoder V4'-V1'. (3) The imagined figure 6 is then fed into visual encoder, and language 'rotate .', could flip the figure upside down into 9. (4) 'it is ?' can elicit the identity of the manipulated figure via MTG, and articulates 'nine' via Broca-PM-M1. (5) the imagined nine can be future manipulated by other language commands such as 'shrink' iteratively. Red arrows indicate that V1 input comes from previously reconstructed imagination.

*action B*', where *condition*, *action A* and *action B* were assigned with one or two random words of vocabulary (Supplementary Methods). As displayed in Fig. 4D, after dlPFC hearing the statement, it started to output query '*condition* ?'; if received the answer 'T', dlPFC output *action A*, otherwise *action B*. After sufficient training, conditional selection in dlPFC could also be generalized to word 'sep' beyond the training vocabulary.

Figure 5D-F demonstrate how HGLP learns the 'trace' and 'write' skills at one trial guided by language. After telling HGLP 'if sep, SAY turn, else .', the sentence was first encoded into 32-byte *lv* by A1-SPT language encoder, then decomposed into word level *lv* by Wernicke module. BA39/40 and MTG did not respond, but dlPFC (Fig. 4D) knew how to handle this 'if then' command by outputting 'sep ?' to query whether pen and digit template would separate. Meanwhile, the digit template was converted into 32-byte *vv* by visual encoder V1-V4, which accompanied with 'sep ?' were received and processed by MTG. As is shown in Fig. 3G-H, MTG predicted the separation by saying 'T' when the pen would leave the digit template, otherwise saying 'F', which were received by dlPFC. If MTG's feedback was 'T', dlPFC would initiate the articulation of 'turn' with the help of Broca, Premotor and M1 (current HGLP has not implemented sensorimotor modules yet), so as to let the external pen to randomly turn its motion direction. In this way, the HGLP traced digit 5 like a child according to the heard language command and viewed digit template (Supplementary Movie S1). Finally, the abstract knowledge *lv* of 'trace is 'if sep, SAY turn" could be saved in BA39/40 for future use. The writing skill can be decomposed as 'imagine a template, trace it'. MTG first imagines a digit (Fig. 2B) which could be taken as a template at V1 and be traced by the verbally controlled external pen. The Movie S2 and Fig. 4F show the writing progress and final result, respectively. This tracing and writing processes demonstrate skill learning guided by language instruction, and later repeated practice will turn it into habitual skill with reinforcement learning implemented in basal ganglia [26-28].

The hippocampus and its surrounding structures are very important for navigation, episode memory, and learning. The brain records the daily attended events as episodic memory into hippocampus [29], and then consolidate the abstract knowledge into neocortex through the engram mechanism [30, 31]. We have not implemented the mechanism into HGLP yet, but simply organizes the training data in the form of a hippocampus-like template, so as to supervised train the above association and dlPFC modules. For example, to train MTG, the hippocampus-like template needs to prepare both input and expected output n×32-byte *lv* and *vv*, where n = 5 indicates that each sentence has five frames of words or phrases. Its input and output contents are generated according to tasks, which are scheduled from easy to hard, just like teaching kids. First train HGLP some simple tasks such as object recognition, and then more complex and integrated tasks, which can take use of previously learned skills flexibly. For any task, we need to arrange the hippocampal template to provide the corresponding input and expected output for each module. Even those modules that are not directly related to the task should learn to give a silent response (Fig. 1C-D). The development of HGLP depends on the arrangement of tasks to be trained. Learning new tasks will inevitably affect the performance of old skills, so it is necessary to occasionally re-experience specific tasks in order to keep the skills and knowledges from being forgotten. So, the contents saved in the hippocampus determine the direction of knowledge development of cortical modules.

HGLP has two routing directions (Fig. 1): the bottom-up route conveys the processed sensory information to high-level modules and the response of association modules to dlPFC's query; the top-down route conveys the verbal attention and query from dlPFC to association modules, and the control signals to sensorimotor decoders for visual imagination and verbal articulation. These enable the HGLP network to handle all kinds of information, interact with the outside world and even self-think flexibly and effectively. Note that, as there are a large number of interconnections among various brain regions, HGLP also supports these interconnections if module A's output modality is the same as module B's input. For example, MTG's or BA39 language output can be fed into Broca.

After learning the above skills, HGLP could perform the human-like thinking process guided by language in Fig. 4G. (1) HGLP first closed its eyes, namely that, no external images were fed into V1 (all the subsequent input images were generated through the imagination process). (2) Hearing 'six .', MTG translated it into the corresponding *vv* that could be reconstructed into an imagined figure by V4'-V1'. (3) The imagined figure 6 is then fed into visual encoder, and language 'rotate .' could flip it upside down into 9 as Fig. 3E. (4) Following the sentence 'it is ?', HGLP could consciously identify the manipulated figure and articulate 'nine' via Broca-PM-M1. (5) Finally, HGLP used 'shrink .' to make the virtual digit smaller, where red arrows indicate that V1 input comes from the previously reconstructed imagination. This experiment demonstrates HGLP could understand the word and sentence by properly manipulating imagination and generating verbal output. The HGLP also forms the iterative thinking process guided by language.

## 3   Discussion and Conclusion

HGLP demonstrates the ability to solve those five questions raised at the beginning. (1) Step by step learning is achieved by proper arrangement of task learning schedule with hippocampal templates. Previously learned skills can be flexibly reused in subsequent tasks. For example, to understand the word meaning, HGLP needs to be able to repeat the heard pronunciation; and word writing task (Fig. 4F) requires the digit imagination and understanding of 'if then' statement. (2) Word meaning is explained with multimodal neuronal activation. Previously, activating digit 2 related visual neurons requires to visualize a digit 2 instance, but now (Fig. 3B) with the verbal word 'two' and the MTG translator, those 2 related visual neurons can be activated. Or we can say verbal 'two' is explained by these visual neurons. (3) Use the virtual world to understand and generate sentences. This virtual world is dynamically built in HGLP's visual autoencoder as Baddeley's VSS and language autoencoder as Baddeley's PL. The imagined visual output can be fed back to the visual encoder and the sentence to be articulated can be fed back to the language encoder, so the processed information is not lost after articulation or imagination. The visual autoencoder is a mental stage, where objects can be created and manipulated, the PL can be viewed as a script to guide the development of a story on the stage, and dlPFC is the director. The mental virtual world is built and maintained in real-time according to the language and visual input from the outside world and the control commands from dlPFC, so HGLP has the ability to interact with the environment through the virtual world model. HGLP can manipulate its mental world according to other people's language to understand their intention, meanwhile, it can generate verbal reply according to the evolution of the virtual world. (4) The thinking process of HGLP is expressed via *vv* and *lv*, which could be visualized and articulated via HGLP's own visual and language decoders. Like human consciousness stream, it is hard to know what you are thinking via any brain measurements, but you can easily express it via your articulation system. The *vv* and *lv* streams have the same characteristics. (5) The 'digit tracing' experiment demonstrates that HGLP has the human-like ability of one trial learning. It understands the rule conveyed by language and knows how to behave according to the rule. This is a goal-directed rapid learning system, a complementary system to the habitual learning system implemented with the reinforcement learning framework.

HGLP also demonstrates some other human-like features. It could identify digit instance at a glance, instead of ranking every item in vocabulary by softmax [32]. Just like people can immediately recognize watermelon without evaluating the probability of being a cheery or car. In this way, HGLP does not need to have a fixed vocabulary size, and can dynamically remember novel items if learned or forget items that are not experienced frequently. Moreover, HGLP is a human-like neural network because it shows similar symptom when special module is lesioned. Fig. 2E demonstrated the Broca aphasia symptom when neurons in the Broca module were silenced. We also know, by removing the hippocampus template, HGLP will behave like patient H.M. [33] who preserved sensorimotor functions, working memory, learned knowledge, etc. but could not form episodic memory nor learning new knowledge. In addition, HGLP also demonstrated a more human-like voluntary attention mechanism guided by language (Fig. 3F), which future allows machine to act according to its won choice mediated by its inner language *lv* to achieve 'free will in machine', that is beyond the capacity of current self-attention mechanism [1-3].

In the future, we will add other human-like brain modules, such as sensorimotor system and basal ganglia system, so that robots can have both goal-directed action output under HGLP language control and non-describable habitual behavior under reinforcement learning control. We may also add a human-like value system including insular, amygdala and OFC, etc. to detect the interoception and voluntarily generate language to guide its own thinking and behavior [34, 35]. We will also divide visual autoencoder into ventral stream and dorsal stream to process object features and spatial information respectively [36]. Some other cortices can also be implemented under the HGLP framework to make it more intelligent and human-like.